\documentclass{article}
\usepackage{emulateapj,pstricks,apjfonts}

\righthead{SARAZIN \& KEMPNER}
\lefthead{NONTHERMAL BREMSSTRAHLUNG IN CLUSTERS OF GALAXIES}

\slugcomment{Astrophysical Journal, in press}

\begin{document}

\title{Nonthermal Bremsstrahlung and Hard X-ray Emission from
Clusters of Galaxies}

\author{Craig L. Sarazin and Joshua C. Kempner}

\affil{Department of Astronomy, University of Virginia,
P.O. Box 3818, Charlottesville, VA 22903-0818;
cls7i@virginia.edu,
jck7k@virginia.edu}

\begin{abstract}
We have calculated nonthermal bremsstrahlung (NTB) models for the hard X-ray
(HXR) tails recently observed by {\it BeppoSAX} in clusters of galaxies.
In these models, the HXR emission is due to suprathermal electrons with
energies of $\sim$10--200 keV.
We consider models in which these transrelativistic suprathermal particles are
the low energy end of a population of electrons which are being
accelerated to high energies by shocks or turbulence
(``accelerating electron'' models).
We also consider a model in which these electrons are the remnant of
an older nonthermal population which is losing energy and rejoining
the thermal distribution as a result of Coulomb interactions
(``cooling electron'' models).
The suprathermal populations are assumed to start at an electron
kinetic energy which is $3 kT$, where $T$ is the temperature of the
thermal intracluster medium (ICM).
The nonthermal bremsstrahlung spectra flatten at low photon energies
because of the lack of low energy nonthermal particles.
The accelerating electron models have HXR spectra which are nearly
power-laws from $\sim$20--100 keV.
However, the spectra are brighter and flatter than given by the
nonrelativistic bremsstrahlung cross-section because of transrelativistic
effects.
The HXR spectrum of the cooling electron model is very flat, and most
of the X-ray emission in the HXR energy range (10-100 keV) actually
arises from electrons with much higher energies ($\sim$100 MeV).
Under the assumption that the suprathermal electrons form part of
a continuous spectrum of electrons including highly relativistic
particles, we have calculated the inverse Compton (IC)
extreme ultraviolet (EUV), HXR, and radio synchrotron emission by the
extensions of the same populations.
For accelerating electron models with power-law momentum spectra 
($N[p] \propto p^{- \mu}$) with $\mu \la 2.7$, which are those expected
from strong shock acceleration, the IC HXR emission exceeds that due
to NTB.
Thus, these models are only of interest if the electron population is
cut-off at some upper energy $\la$1 GeV.
Similarly, flat spectrum accelerating electron models
produce more radio synchrotron emission than is observed from clusters if
the ICM magnetic field is $B \ga 1$ $\mu$G.
The cooling electron model produces vastly too much EUV emission as
compared to the observations of clusters.
We have compared these NTB models to the observed HXR tails in
Coma and Abell~2199.
The NTB models require a nonthermal electron population which contains
about 3\% of the number of electrons in the thermal ICM.
If the suprathermal electron population is cut-off at some energy
above 100 keV, then the models can easily fit the observed HXR fluxes
and spectral indices in both clusters.
For accelerating electron models without a cutoff, the electron spectrum
must be rather steep $\ga 2.9$ to avoid producing too much IC HXR
emission.
The model HXR spectra are then rather steep, but marginally consistent
with observations of the HXR spectrum in Abell~2199 and Coma or the radio
spectrum in Coma.
These models can account for the HXR and radio properties of these
two clusters, but do not produce enough EUV emission.
\end{abstract}


\section{Introduction} \label{sec:intro}

Recent observations with {\it BeppoSAX} have detected hard
X-ray (HXR) tails in the X-ray spectra of the Coma cluster
(Fusco-Femiano et al.\ 1999)
and Abell~2199
(Kaastra, Bleeker, \& Mewe 1998;
Kaastra et al.\ 1999).
These tails, which have been fit by power-law spectra, are an
excess to the thermal X-ray emission from the hot intracluster medium
(ICM).
In Coma, which has a diffuse radio halo
(e.g., Deiss et al.\ 1997), there have been many previous attempts
to detect nonthermal HXR emission produced by inverse Compton
(IC) scattering of cosmic microwave background (CMB) photons
by the radio-emitting relativistic electrons
(e.g., Rephaeli et al.\ 1994).
Comparison of the radio synchrotron and IC HXR emission allow one
to determine the magnetic field in the ICM
(e.g., Rephaeli 1979);
in Coma, Fusco-Femiano et al.\ (1999) find $B \approx 0.16$ $\mu$G.

All in all, IC scattering is probably the most attractive model
for the HXR emission in clusters, but it is not without problems.
First, the magnetic field in Coma 
is smaller than that determined from
Faraday rotation toward individual radio galaxies
($\sim$6 $\mu$G; Feretti et al.\ 1995)
or equipartition in the diffuse radio halo
($\sim$0.4 $\mu$G; Giovannini et al.\ 1993;
En{\ss}lin \& Biermann 1998).
One way of describing this discrepancy would be to say that the
observed HXR emission is stronger than expected for models of the
radio source with the larger magnetic fields.

The situation in Abell~2199 is more extreme.
This cluster lacks any extended diffuse radio emission
(Kempner \& Sarazin 1999),
and thus no IC HXR emission was expected.
The detection of this HXR emission implies a very weak ICM magnetic
field of
$\la0.07$ $\mu$G
if the HXR emission is IC
(Kempner \& Sarazin 1999).

The weak ICM magnetic field in Coma and very restrictive limit in
Abell~2199 could be increased if some other mechanism contributed
to the HXR emission in clusters.
One suggestion is that all or part of this emission might be nonthermal
bremsstrahlung (NTB) from suprathermal electrons with energies
of $\sim$ 10 -- 200 keV
(Kaastra et al.\ 1998;
En{\ss}lin et al.\ 1999).
These nonthermal electrons would form a population in excess of
the normal thermal gas which is the bulk of the ICM.
Perhaps the most natural explanation of this suprathermal population
would be that they are particles which are currently being accelerating
to higher energies, either by shocks or turbulence in the ICM.

Nonthermal bremsstrahlung has also been invoked as a possible explanation
for hard X-ray tails in the spectra of supernova remnants
(Skibo, Ramaty, \& Purcell 1996;
Baring et al.\ 1999),
such as
SN~1006
(Koyama et al.\ 1995),
Cas~A
(Allen et al.\ 1997),
IC~443
(Keohane et al.\ 1997),
RXJ1713.7-3946
(Koyama et al.\ 1997),
and
RXJ0852.0-4622
(Allen, Markwardt, \& Petre 1999).
Similar arguments have been made for the hard X-ray emission seen from
the Galactic ridge
(Kaneda et al.\ 1997;
Valinia \& Marshall 1998).

In this paper, we calculate models for the nonthermal bremsstrahlung
in clusters of galaxies.
A variety of models for the suprathermal electrons are developed
in \S~\ref{sec:particles}.
The nonthermal bremsstrahlung fluxes and spectra are
calculated in \S~\ref{sec:bremss}.
The nonthermal electron populations required might also extend
to much higher energies, at which the electrons would be fully
relativistic.
In \S~\ref{sec:harder},
the resulting IC HXR emission, IC extreme ultraviolent (EUV) emission,
and diffuse radio emission are derived.
The models are compared to the observations of Coma and Abell~2199
in \S~\ref{sec:observe}.
Finally, our conclusions are presented in \S~\ref{sec:conclusion}.

\section{Nonthermal Particle Populations} \label{sec:particles}

We will assume that most of the intracluster electrons are part of a
thermal distribution with a temperature $T$.
Let $N_{th}^{tot}$ be the total number of thermal electrons in the
cluster.
For the specific numerical models we give, we will assume that
$k T = 7$ keV and 
$N_{th}^{tot} = 10^{71}$,
which corresponds approximately to a total
thermal gas mass of $10^{14} \, M_\odot$.
Here, we will be interested in nonthermal electrons, which we take to be
a higher energy population in excess of the thermal population.
In this paper, we will ignore any spatial variations in the spectrum
of nonthermal electrons, and concentrate on the integrated population
throughout the cluster.
We will represent the distribution of nonthermal particles as a function
of their momentum.
Let $N ( P ) d P$ be the total number of nonthermal electrons with 
momenta in the range $P$ to $P + dP$.
We will use the normalized electron momentum defined by
$p \equiv P/(m_e c)$.

We will assume that the nonthermal population consists of electrons
with kinetic energies $E > 3 k T$.
Let $p_l$ be the normalized electron momentum which corresponds to this
kinetic energy,
\begin{equation} \label{eq:pl}
p_l = \left[ \left( 1 + \frac{3 k T}{m_e c^2} \right)^2 - 1 \right]^{1/2}
\, .
\end{equation}
For the numerical models with $kT = 7$ keV, $p_l = 0.2896$.

We will normalize the nonthermal electron models such that the total
number of nonthermal particles is 1\% of the
thermal populations of electrons
($N_{nt}^{tot} = 10^{69}$ electrons in the numerical models).
The models we consider are summarized in Table~\ref{tab:models}.

%
%
\begin{table*}[htb]
\tabcaption{\hfil Models for the Nonthermal Bremsstrahlung Hard X-ray Emission
\label{tab:models} \hfil}
\begin{center}
\begin{tabular}{lccccccc}
\tableline
\tableline
&&&&&&&\cr
&$N_{nt}^{tot}$&$N_o$&$\dot{N}$&$L_\epsilon$ (30 keV)&$L_{HXR}$&
$L_\epsilon^{fit}$ (20 keV)&$\alpha^{fit}$\cr
Model&($10^{69}$)&($10^{68}$)&($10^{56}$ s$^{-1}$)&
($10^{41}$ ergs s$^{-1}$ keV$^{-1}$)&($10^{43}$ ergs s$^{-1}$)&
($10^{41}$ ergs s$^{-1}$ keV$^{-1}$)&\cr
\tableline
Power-law, $\mu =2.0$&1&2.896\phn&$>$0.51&7.33&7.38&9.79&$-$0.708\cr
Power-law, $\mu =2.3$&1&2.596\phn&$>$0.66&6.14&5.59&8.87&$-$0.894\cr
Power-law, $\mu =2.6$&1&2.203\phn&$>$0.82&5.25&4.38&8.23&$-$1.091\cr
Power-law, $\mu =3.0$&1&1.677\phn&$>$1.02&4.41&3.36&7.69&$-$1.355\cr
Power-law, $\mu =3.5$&1&1.128\phn&$>$1.28&3.67&2.63&7.31&$-$1.678\cr
Power-law, $\mu =4.0$&1&0.7286   &$>$1.53&3.14&2.20&7.09&$-$1.988\cr
Nonlinear, $p_c =0.3$&1&2.325\phn&$>$0.78&6.23&6.02&8.85&$-$0.835\cr
Nonlinear, $p_c =0.5$&1&1.816\phn&$>$1.02&5.24&4.80&8.07&$-$1.007\cr
Nonlinear, $p_c =1.0$&1&1.164\phn&$>$1.32&3.98&3.24&7.26&$-$1.404\cr
Cooling Electrons    &1&0.0168   &$-$$3.8 \times 10^{-5}$&1.50&2.37&1.60&$-$0.161\cr
\tableline
\end{tabular}
\end{center}
\end{table*}

\subsection{Accelerating Populations} \label{sec:particles_accel}

We consider a number of models of the transrelativistic nonthermal electron
distribution which might result from particles currently being
accelerated out of the thermal population.
First, we assume that the electrons result from ongoing first-order
Fermi shock acceleration.
If the accelerating particles are treated as test particles, kinetic
theory indicates that the particle spectrum is a power-law in the momentum
(Axford, Leer, \& Skadron 1977;
Bell 1978a,b;
Blandford \& Ostriker 1978):
\begin{equation} \label{eq:powerlaw}
N ( p ) = N_o p^{- \mu} \, \qquad p \ge p_l
\, .
\end{equation}
We will refer to these models as ``power-law'' (PL) models.
In this and the other models below, the parameter $N_o$ gives the 
population at $ p = 1$.
For shock acceleration, the exponent $\mu = ( r + 2 ) / ( r - 1 )$,
where $r$ is the shock compression.
In supernova remnants in our Galaxy, the radio spectra suggest that
the acceleration produces electrons with $\mu = 2.0$--2.6.
Thus, we will consider models with values of $\mu = 2.0$, 2.3, and 2.6
(Table~\ref{tab:models}).
More complicated shock geometries (non-plane parallel shocks or non-normal
magnetic field orientations) can result in more complicated
particle spectra
(e.g., Jokipii 1987).

There are a number of arguments that suggest that steeper power-law
spectra might occur in cluster of galaxies.
First, if the acceleration occurs in intracluster shocks due to cluster
mergers, the compression can be lower than in supernova remnants.
The gas in the merging subclusters is already quite hot, and the typical
Mach numbers of the shocks are only several, rather than being very large.
For example, Markevitch, Sarazin, \& Vikhlinin (1999) recently analyzed
the temperature structure in three clusters which are currently undergoing
fairly major mergers.
In the Cygnus A cluster, which had the simplest geometry and analysis,
the compression associated with the shock was $r \approx 2.2$.
In Abell 3667, the compression was $r \approx 2.5$.
These shock compressions would imply $\mu \approx 3$--3.5.
Second, it is also possible that turbulent acceleration plays a significant
role in clusters of galaxies
(e.g., Eilek \& Weatherall 1999).
For example, the centrally located, diffuse ``radio halos'' in clusters might
be due to synchrotron emission from electrons produced by turbulent
acceleration, while the more localized ``radio relics'' were associated
with shock acceleration.
Turbulent acceleration, which is generally a second-order Fermi process,
might produce a steeper particle acceleration than shock acceleration.
Thus, we have also calculated models with power-law electron momentum
distributions and with steeper spectra ($\mu = 3$, 3.5 and 4 in
Table~\ref{tab:models}).

The energy associated with the acceleration of relativistic particles
in shocks can have significant effects on the shock structure
(e.g., Ellison, Jones, \& Reynolds 1990).
The pressure of the accelerating particles affects the upstream plasma,
and a precursor shock may form.
At same time, energy losses to escaping accelerating particles increase
the overall shock compression
(Eichler 1984).
Although most of the energy may be associated with accelerating ions,
the changes to the shock structure can affect the acceleration of
electrons as well.
The result is to steepen the electron spectrum at low momenta
(Baring et al.\ 1999).
For typical nonlinear shock structure, the electron spectrum steepens
from being proportional to $p^{-2}$ at high energies to being proportional
to $p^{-4}$ at low energies.
As our second model, we adopt a simple parameterized electron spectrum based
on these nonlinear shock models, with
\begin{equation} \label{eq:nonlinear}
N ( p ) = N_o p^{-2} \,
\frac{\left[ 1 + \left( \frac{p_c}{p} \right)^2 \right]}
{\left( 1 +  p_c^2 \right)} \, \qquad p \ge p_l \, .
\end{equation}
The value of $p_c$ should depend on the diffusion of electrons
in the shocks and on the details of the shock structure.
The appropriate averages of these properties are unknown in clusters.
Instead, we will adopt values of $p_c = 0.3$, 0.5, and 1, which imply 
that the electron spectrum steepens at the low energy end of the range
of interest in clusters.
For much lower values of $p_c$, equation~(\ref{eq:nonlinear})
reduces to equation~(\ref{eq:powerlaw}) for electron momenta of interest
for nonthermal bremsstrahlung.
We will refer to these models as ``nonlinear'' (NL) models.

\subsection{Cooling Electrons from an Old Population}
\label{sec:particles_cool}

As an alternative to models in which the nonthermal particles are
currently being accelerated, we will also consider a model in which
these particles arise from an older population of relativistic electrons
(e.g., Kardashev 1962).
Recently, EUV/soft X-ray emission has been detected from clusters which
might be due to inverse Compton (IC) emission from a populations of
relativistic electrons with energies of $\sim$150 MeV
(Sarazin \& Lieu 1998).
These particles have lifetimes in clusters which are $>$$10^9$ yr,
and should persist long after the process which accelerated the relativistic
particles has ceased.
The lower energy nonthermal particles will lose energy more rapidly due
mainly to Coulomb losses to the thermal plasma. 
These energy losses could provide a supply of low energy nonthermal
particles for nonthermal bremsstrahlung.
We will refer to these models as ``cooling electron'' models.

The timescales for Coulomb losses are short for the low energy nonthermal
particles we consider here.
As a result, the cooling electrons are expected to be in steady-state
(Sarazin 1999, \S~3.1.6).
The rate of energy loss due to Coulomb interactions is given by
(e.g., Rephaeli 1979)
\begin{equation} \label{eq:coulomb}
\frac{d E}{d t} = - \, \frac{4 \pi n_e e^4}{m_e c} \,
\left( \frac{c}{v} \right) \, 
\ln \left[ 1.12 \gamma^{1/2} \left( \frac{v}{c} \right)^2 \,
\frac{m_e c^2}{\hbar \omega_p} \right] \, ,
\end{equation}
where $v$ and $\gamma$ are the velocity and Lorentz factor of the
nonthermal electron, $n_e$ is the thermal electron number density,
and $\omega_p = ( 4 \pi n_e e^2 / m_e )^{1/2}$ is the 
plasma frequency in the thermal gas.
The Coulomb logarithm term used here is approximate;
it applies in either the subrelativistic quantum limit ($0.01 \ll (v/c) \ll 1$)
or in the ultrarelativistic limit.
However, the argument of the logarithm is generally very large, and the 
variations due to $v$ and $\gamma$ are quite slow.
The only major variation of the loss rate with energy in
momentum comes from the factor $( c / v )$ outside of the logarithm.
For the purpose of constructing this simple model, we will ignore the
variation in the Coulomb logarithm, which only affects the solution at
the level of a few percent.

Then, if one follows the same steady-state argument for the low energy
population given in Sarazin (1999, \S~3.1.6), the momentum distribution
of the cooling electrons is found to be
\begin{equation} \label{eq:cool}
N ( p ) = 2 N_o \, \frac{p^2}{1 + p^2} \, \qquad p_l \le p \le p_u \, .
\end{equation}
The maximum in the loss time of electrons in a cluster occurs for
$E \sim 150$ MeV or $p \sim 300$.
At energies which are higher than this, electrons lose energy
rapidly due to IC and synchrotron radiation.
Thus, we will assume an upper limit of $p_u = 300$ for this cooling
electron distribution.

\centerline{\null}
\vskip2.55truein
\includegraphics{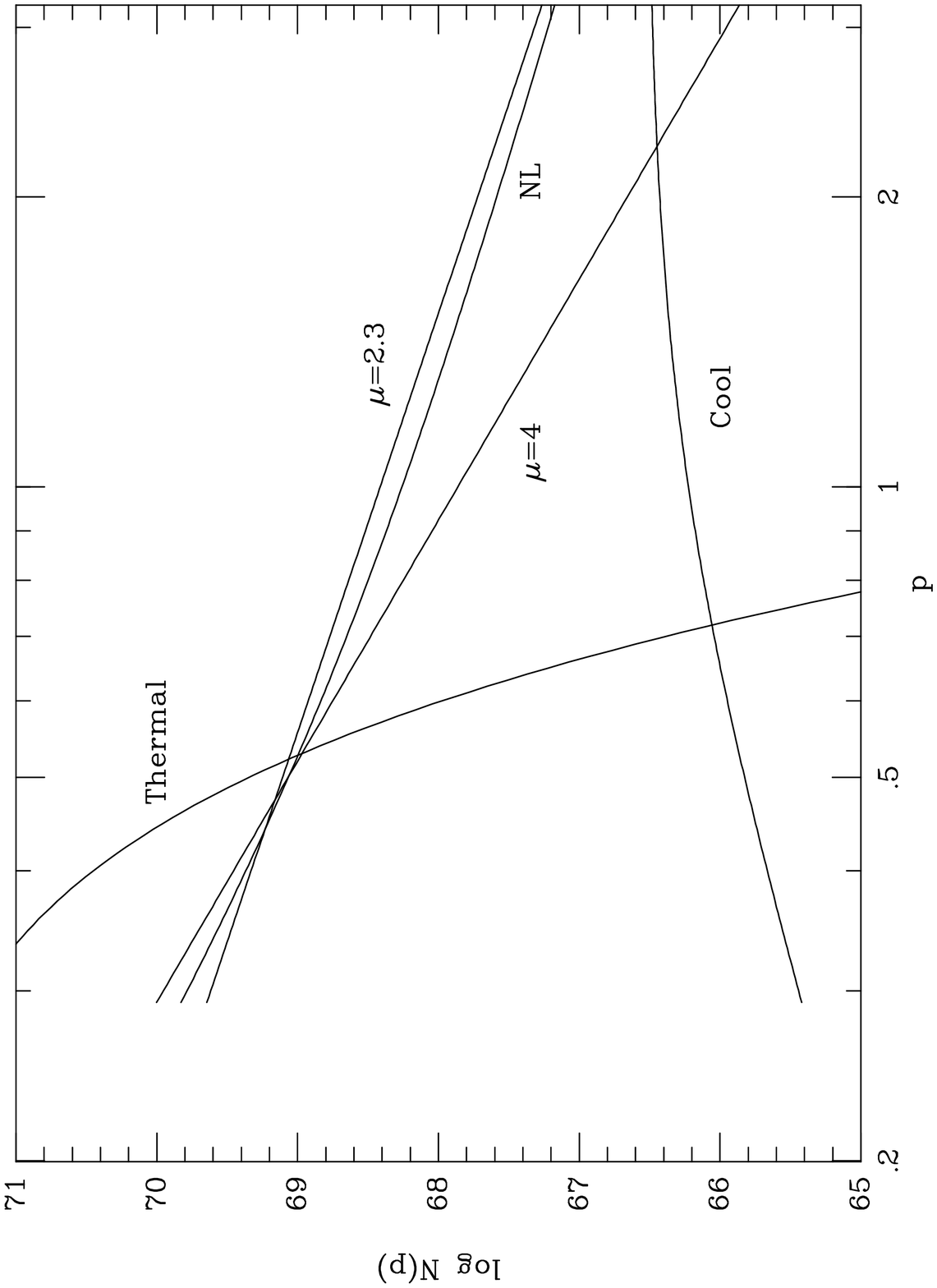}
\figcaption{Four models for the nonthermal electron population in
clusters, as a function of the normalized momentum $p \equiv P / m_e c$.
There are two models with power-law momentum distributions
(eq.~[\protect\ref{eq:powerlaw}]) with
$\mu = 2.3$ and $\mu = 4$,
a nonlinear shock model with $p_c = 0.5$
(``NL,'' eq.~[\protect\ref{eq:nonlinear}]),
and the cooling electron model
(``Cool,'' eq.~[\protect\ref{eq:cool}]).
The thermal distribution for $kT = 7$ keV is shown for comparison.
\label{fig:pop_p}}

\vskip0.2truein

\subsection{Resulting Nonthermal Electron Populations}
\label{sec:particles_result}

Four examples of the resulting nonthermal populations are shown
in Figure~\ref{fig:pop_p}.
The models are
a power-law momentum distribution with $\mu = 2.3$
(eq.~[\ref{eq:powerlaw}]),
a power-law momentum distribution with $\mu = 4$,
a nonlinear shock acceleration model with $p_c = 0.5$
(``NL,'' eq.~[\ref{eq:nonlinear}]),
and a cooling electron population
(``Cool,'' eq.~[\ref{eq:cool}]).
For comparison, the high energy tail of the fiducial thermal population,
calculated using the transrelativistic Maxwellian distribution,
is shown as the curve labeled ``Thermal.''
The thermal distribution is for $kT = 7$ keV and $N_{th}^{tot} = 10^{71}$.
The nonthermal distribution are normalized as discussed above, and
the normalization parameter $N_o$ is given in Table~\ref{tab:models}.
In Figure~\ref{fig:pop_tot_e}, we
show the resulting total thermal and nonthermal distributions in these
same four models, plotted as energy distribution functions.
The dashed curve is for a purely thermal distribution.

As discussed above, the nonthermal electron population was normalized by
comparing the number of electrons to the corresponding number in the thermal
population.
The flux of electrons through the subrelativistic, nonthermal population
is also of interest.
Let $\dot{N}$ be the number of electrons passing through this population
per unit time, with a positive sign implying the electrons are moving to
higher energies.
For the cooling electron population, this number flux is a constant and
is determined by the rate of Coulomb energy losses (eq.~[\ref{eq:coulomb}]).
For accelerating electrons, the flux depends on the time scale for electron
acceleration, which is uncertain in clusters.
However, efficient acceleration probably requires that the rate of
acceleration exceed the rate of energy losses.
Based on this argument, the lower limit to the number flux of accelerating
particles was calculated using equation~(\ref{eq:coulomb}).
The lower limit is largest at the lower limit momentum $p_l$, so this
is where it was evaluated.
The average thermal electron density in the cluster was assumed to be
$n_e = 0.001$ cm$^{-3}$.
These lower limits on $\dot{N}$ for accelerating models and the value
for the cooling model are given in Table~\ref{tab:models}.

\centerline{\null}
\vskip2.55truein
\includegraphics{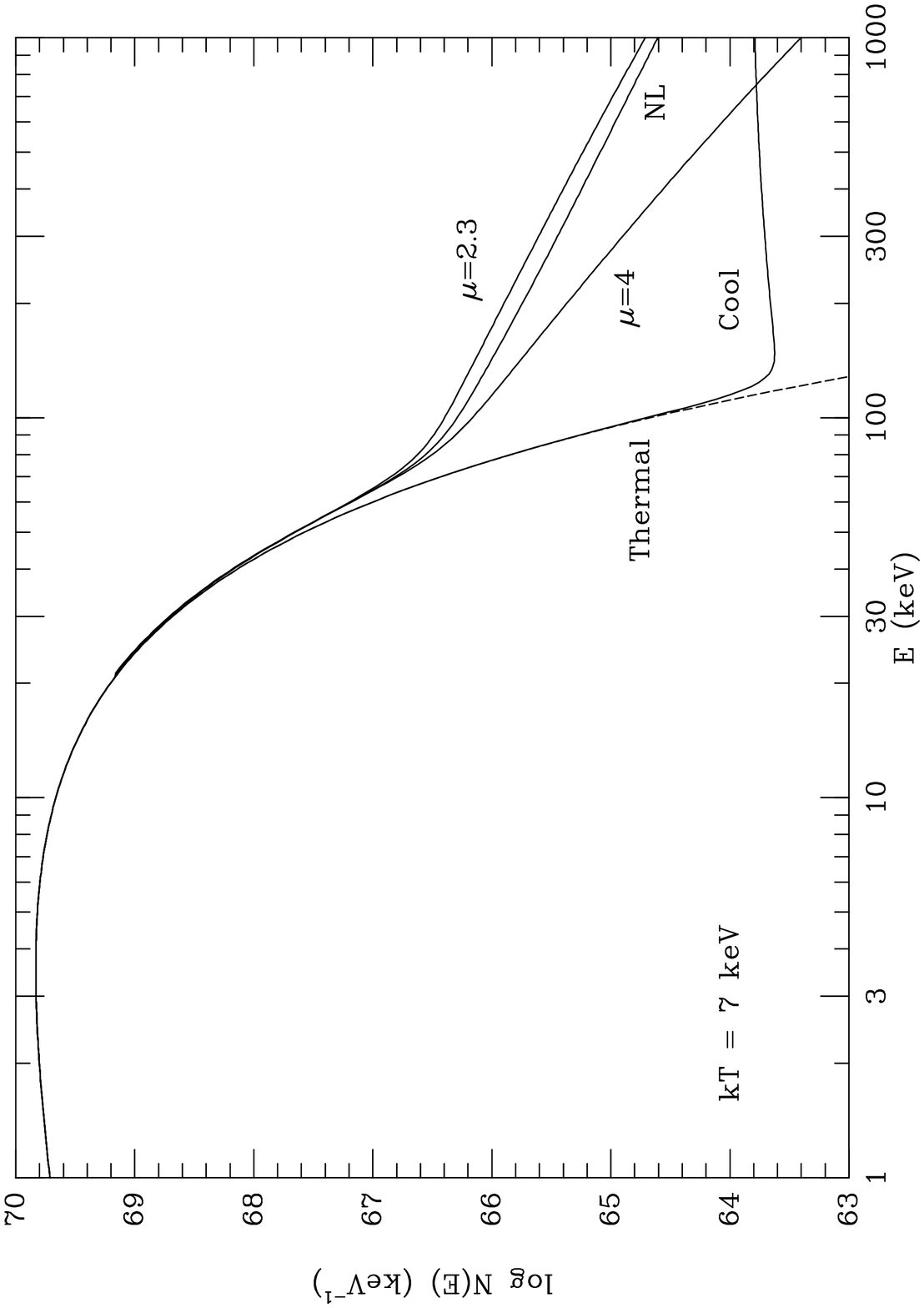}
\figcaption{The resulting total (thermal plus nonthermal) energy
distributions for the same four nonthermal models as shown in
Figure~\protect\ref{fig:pop_tot_e}.
The distributions are shown as a function of the electron kinetic
energy $E$.
For comparison, the dashed line shows the thermal distribution alone.
\label{fig:pop_tot_e}}

\vskip0.2truein

In the accelerating models, the lower limits on the flux of
nonthermal electrons are
$\dot{N} \ga 0.5 \times 10^{56} ( n_e / 0.001$ cm$^{-3}$) s$^{-1}$.
The total number of thermal electrons is $N_{th}^{tot} = 10^{71}$.
Thus, such acceleration could only continue for a time scale of
$\la 6 \times 10^7 ( n_e / 0.001$ cm$^{-3}$)$^{-1}$ yr before it would
significantly deplete the thermal population.

In the cooling electron models, the flux is from higher to lower energies,
so it is the reservoir of relativistic electrons which is being depleted
(see \S~\ref{sec:harder} below).
In these models, we are assuming that the total number of nonthermal
electrons is 1\% of the thermal population or $N_{nt}^{tot} = 10^{69}$.
The flux of particles to low energies would deplete the nonthermal
population in $\sim 8 \times 10^9$ yr, which is about the maximum loss
time scale for electrons in a cluster, which occurs at
$p \approx \gamma \sim 300$
(Sarazin 1999).

\section{Bremsstrahlung Emission} \label{sec:bremss}

Let $L_\epsilon d \epsilon$ be the luminosity of nonthermal bremsstrahlung
emitted at photon energies 
from $\epsilon$ to $\epsilon + d \epsilon$
(where $\epsilon = h \nu$, and $\nu$ is the photon frequency).
The emission is given by the integral
\begin{equation} \label{eq:emission}
L_\epsilon = \epsilon \, \int_{p_l}^\infty N ( p ) \, d p \,
v ( p ) \, \sum_Z n_Z \frac{d \sigma ( p , \epsilon , Z )}{d \epsilon}
\, ,
\end{equation}
where $v( p ) = c p / ( p^2 + 1 )^{1/2}$ is the velocity
of an electron with a normalized momentum $p$,
$Z$ denotes the charge of the various thermal particles in the plasma
(protons, electrons, helium nuclei, etc.),
$n_Z$ gives the average number densities of these thermal particles,
and
$[ d \sigma ( p , \epsilon , Z ) / d \epsilon ] d \epsilon$ gives the 
cross-section for emitting a photon with an energy in the range
$\epsilon$ to $\epsilon + d \epsilon$ during a collision between a nonthermal
electron with normalized momentum $p$ and a thermal particle $Z$.

\centerline{\null}
\vskip2.55truein
\includegraphics{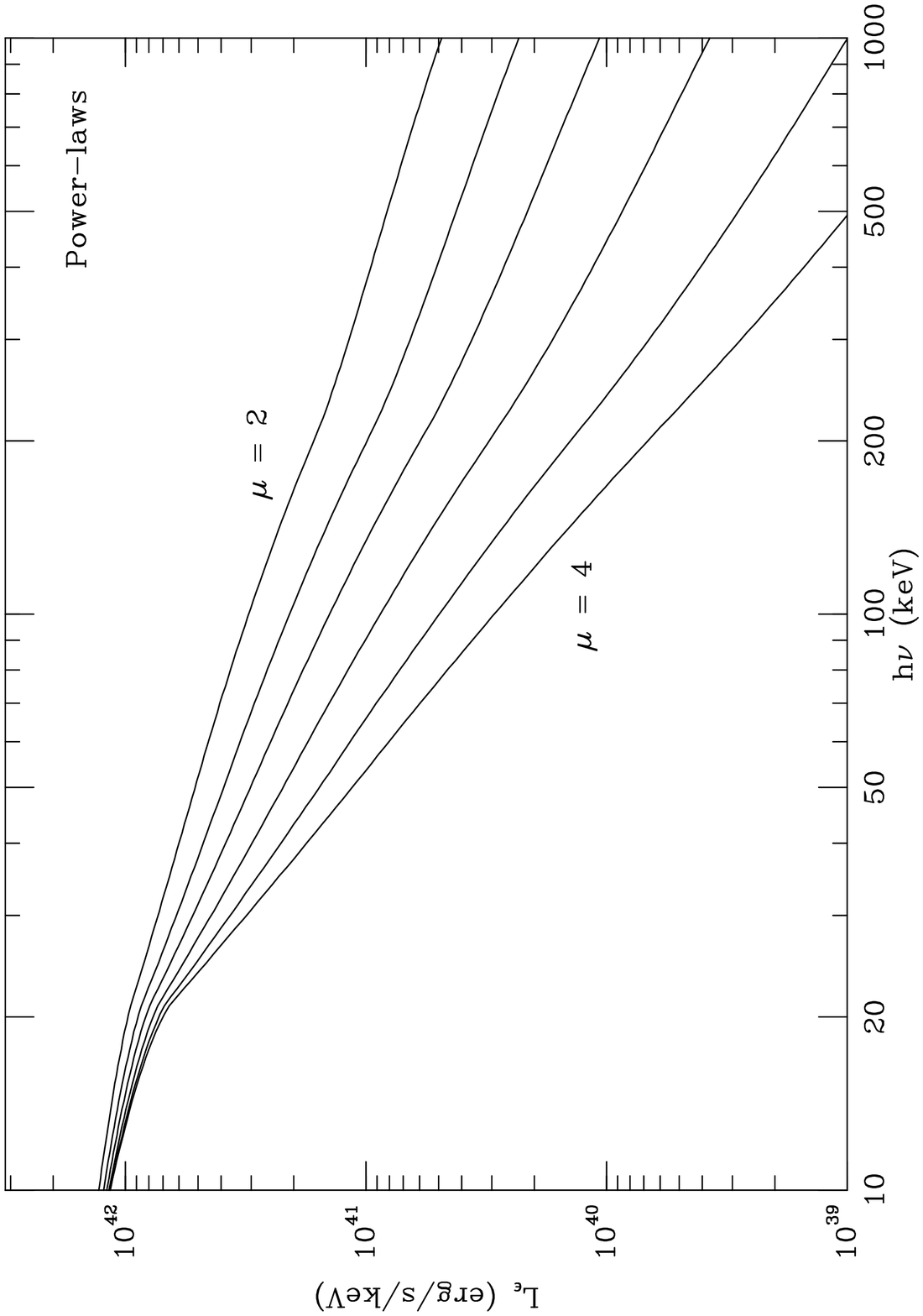}
\figcaption{The nonthermal bremsstrahlung hard X-ray emission of the
models with power-law momentum distributions
(eq.~[\protect\ref{eq:powerlaw}]), with values of the exponent
$\mu = 2$, 2.3, 2.6, 3, 3.5, and 4 (upper to lower).
The emitted spectrum is given as a function photon energy.
\label{fig:ntbrem_pl}}

\vskip0.2truein

For collisions between nonthermal electrons and thermal positive ions,
we use the completely unscreened cross-sections appropriate to a diffuse
ionized thermal plasma
(e.g., Blumenthal \& Gould 1970).
We use the standard nonrelativistic Born approximation expressions 
at low energies
(Koch \& Motz 1959).
Coulomb effects are included by applying the Elwert factor.
At extremely relativistic electron energies,
we use the unscreened relativistic expressions from
Blumenthal \& Gould (1970).
At intermediate energies, we use the transrelativistic expressions given
by Koch \& Motz (1959) and Haug (1997).
At the upper limit for the photon energy $\epsilon \rightarrow \gamma - 1$,
we use the Fano-Sauter approximation
(Koch \& Motz 1959).
We compared our cross-sections to those given by the more general
GALPROP code (Strong \& Moskalenko 1998) kindly provided by Andrew
Strong, and found them to be in excellent agreement.

For the general numerical models,
we assume an average thermal electron density of $n_e = 0.001$ cm$^{-3}$.
We include hydrogen ions, helium ions, and the common heavy element ions.
The abundance of helium was taken to be 9.77\% of that of hydrogen by
number, while the abundances of the heavier elements were one half
of the solar values in
Anders \& Grevesse (1989).
We also include electron-electron bremsstrahlung (Haug 1998), but it
does not contribute very significantly to the emission in the models we
consider here.

\centerline{\null}
\vskip2.55truein
\includegraphics{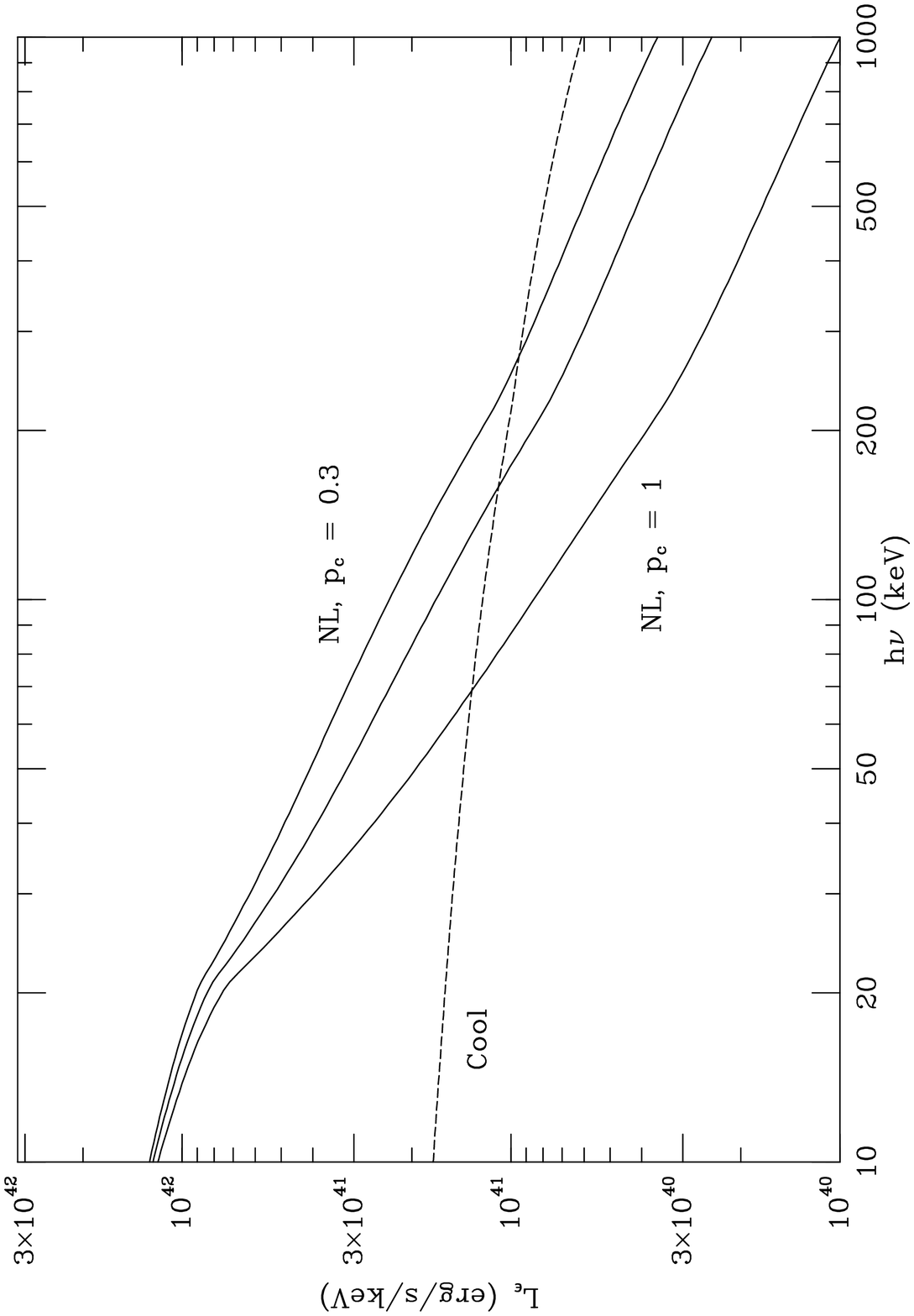}
\figcaption{The nonthermal bremsstrahlung hard X-ray emission for
nonlinear shock models (solid curves, eq.~[\protect\ref{eq:nonlinear}])
and the cooling electron models
(dashed curve, eq.~[\protect\ref{eq:cool}]).
The three nonlinear shock models have characteristic momenta of
$p_c = 0.3$, 0.5, and 1 (upper to lower).
\label{fig:ntbrem_2}}

\vskip0.2truein

Some of the properties of the nonthermal hard X-ray emission of the models
are listed in Table~\ref{tab:models}.
The quantity $L_\epsilon$ (30 keV) gives the nonthermal bremsstrahlung
emission from the models at a photon energy of 30 keV.
The total luminosity for 10 to 200 keV is listed as $L_{HXR}$ in the
table.

The resulting nonthermal bremsstrahlung hard X-ray spectra of the models
are shown in Figures~\ref{fig:ntbrem_pl} and \ref{fig:ntbrem_2}.
The spectra produced by the accelerating electron models show some structure
in this region, which results from the lower energy cutoff to the electron
spectrum and to the transrelativistic regime for the bremsstrahlung
cross-section.
At very low photon energies ($\epsilon \la 20$ keV), these spectra
flatten.
This is the region of the spectrum where the photon energy is less
than the initial kinetic energies of any of the nonthermal electrons
($\epsilon < [ ( 1 + p_l^2 )^{-1/2} -1 ] m_e c^2 = 21$ keV, where the value
applies to the numerical models).
The nonrelativistic Bethe-Heitler bremsstrahlung
cross-section (Heitler 1954) is
\begin{equation} \label{eq:bh}
\frac{d \sigma ( p , \epsilon , Z )}{d \epsilon} =
\frac{32 \pi}{3} \, \frac{e^6}{m_e^2 c^5 h} \, \frac{Z^2}{p_i^2 \epsilon} \,
\ln \left( \frac{p_i + p_f}{p_i - p_f} \right)
\, ,
\end{equation}
where $p_i$ and $p_f$ are the initial and final values of the normalized
electron momentum.
If one considers the limit where the photon energy is much less
than the initial electron energy of an electron, the single-particle
non-relativistic bremsstrahlung spectrum is
\begin{equation} \label{eq:low_e_nonrel}
L_\epsilon = \frac{32 \pi}{3} \, \frac{e^6}{m_e^2 c^4 h} \,
\left( \sum n_Z Z^2 \right) \, \frac{1}{p} \,
\ln \left( \frac{ 2 p^2 m_e c^2}{\epsilon } \right)
\, .
\end{equation}
Thus, the nonthermal bremsstrahlung spectrum only increases as the
logarithm of $(1/\epsilon )$ as $\epsilon$ decreases at photon
energies below the lowest kinetic energy of nonthermal particles.
The same result is found if the relativistic cross-section is 
used (see eq.~[\ref{eq:rel}] below).

It is conventional to fit nonthermal spectra by power-laws, and hard
X-ray tails in clusters have been fit using these functions, under the
assumption that the emission is IC radiation.
We have fit our model spectrum using the power-law form,
\begin{equation} \label{eq:powerfit}
L_\epsilon = L_\epsilon^{fit} (20~{\rm keV}) \,
\left( \frac{ \epsilon }{20~{\rm keV}} \right)^{\alpha^{fit}}
\, ,
\end{equation}
for photon energies of 20 -- 100 keV.
The fits were done by minimizing $\chi^2$, assuming that the variances in
the fluxes were proportional to the number of photons emitted.
The resulting values of $L_\epsilon^{fit}$ (20 keV) and $\alpha^{fit}$
are listed in Table~\ref{tab:models}.

The nonrelativistic bremsstrahlung spectrum of a model with a steeply
declining power-law momentum distribution (eq.~[\ref{eq:powerlaw}])
is a power-law if
$p_l^2 < 2 \epsilon / ( m_e c^2 ) \ll 1 $.
Using the nonrelativistic Bethe-Heitler cross-section (eq.~[\ref{eq:bh}]),
the spectrum of nonrelativistic nonthermal bremsstrahlung is
\begin{eqnarray}
L_\epsilon & = & \frac{32 \pi^{3/2}}{3} \, \frac{e^6}{m_e^2 c^4 h} \,
\left[ \frac{\Gamma( \mu / 2)}{\mu \Gamma( \mu / 2 + 1/2 )} \right]
\nonumber \\
& & \quad \times \left( \sum n_Z Z^2 \right) \,
\, N_o \, \left( \frac{ m_e c^2 }{2 \epsilon } \right)^{\mu / 2}
\, . \label{eq:nt_nonrel}
\end{eqnarray}
where $\Gamma$ is the gamma function.
In Figure~\ref{fig:powerlaw}, the nonrelativistic power-law bremsstrahlung
spectra given by this equation are compared to the detailed calculations
using the full cross-section for our six models with power-law momentum
distributions.
Except for the steepest exponent $\mu=4$,
the detailed spectra are brighter than the nonrelativistic approximation
by factors which range up to about two.
The detailed spectral slopes are also up to 0.3 flatter.
The nonrelativistic approximation predicts
$\alpha = - \mu / 2 = -1$, -1.15, -1.3, -1.5, -1.75, and -2
for $\mu = 2$, 2.3, 2.6, 3, 3.5 and 4.
whereas the best-fit values are
-0.71, -0.89, -1.09, -1.36, -1.68, and -1.99
(Table~\ref{tab:models}).
Both of these differences are greater for the models with smaller $\mu$,
in which relativistic electrons make a larger contribution.
The power-law approximation for the steepest model with $\mu = 4$
works quite well.

\centerline{\null}
\vskip2.55truein
\includegraphics{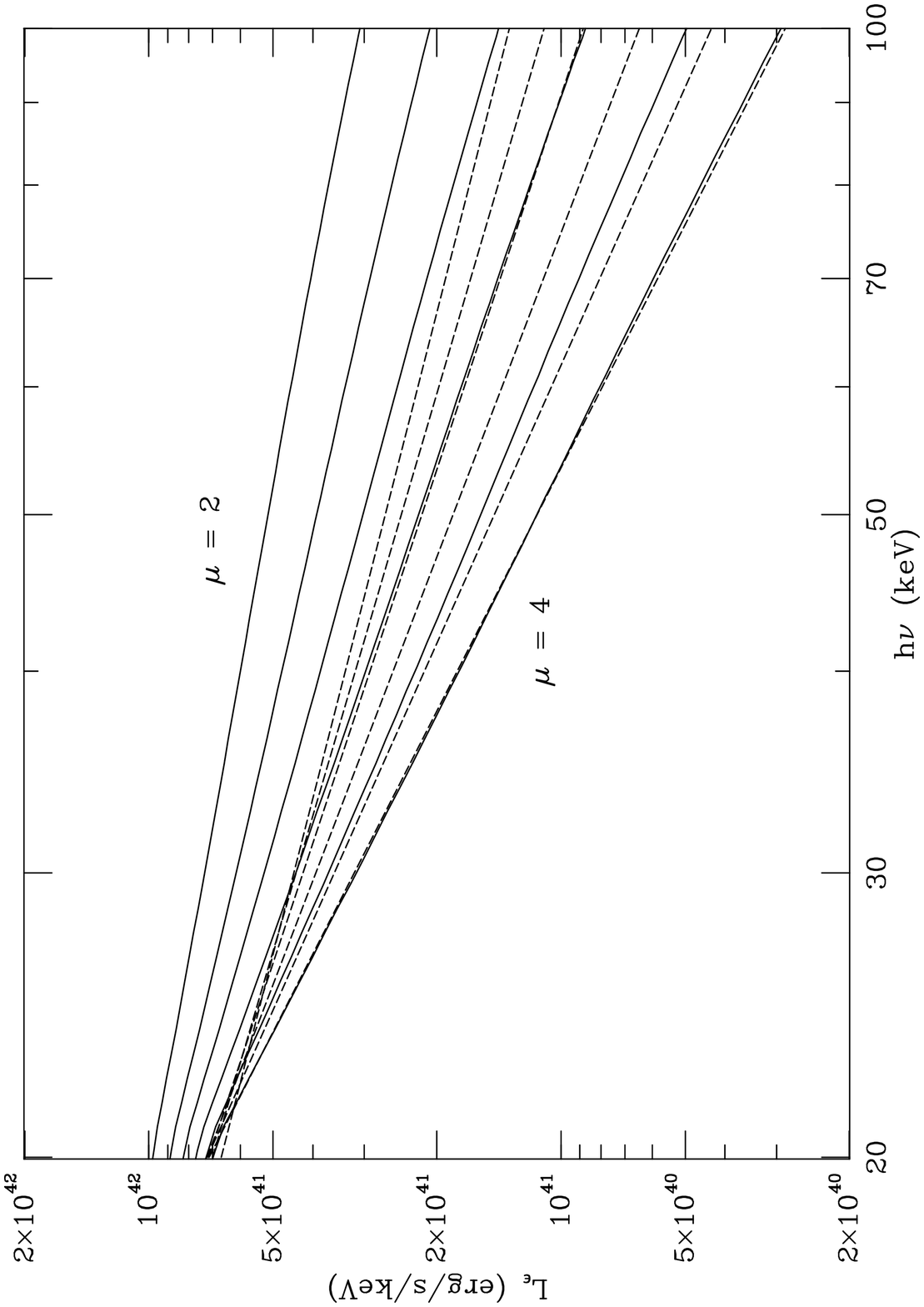}
\figcaption{The detailed nonthermal bremsstrahlung hard X-ray spectra
(solid curves) of the models with power-law electron momentum distributions
(eq.~[\ref{eq:powerlaw}]) are compared to the simple nonrelativistic
power-law approximation to the spectrum
(dashed lines, eq.~[\protect\ref{eq:nt_nonrel}]).
The power-law exponents for the electron momentum are
$\mu = 2$, 2.3, 2.6, 3, 3.5, and 4 (upper to lower).
\label{fig:powerlaw}}

\vskip0.2truein

These differences result from the fact that the low frequency limit
of the relativistic bremsstrahlung cross-section
(Blumenthal \& Gould 1970; eq.~[\ref{eq:rel}] below)
is larger (by a factor of $\sim$$p^2$)
than the nonrelativistic cross-section
(eq.~[\ref{eq:bh}]).
Similar results apply to the nonlinear shock model, since
this model consists essentially of two declining power-law,
$N(p) \propto p^{-4}$ at small momenta and
$N(p) \propto p^{-2}$ at large momenta
(eq.~[\ref{eq:nonlinear}]).

The bremsstrahlung spectrum for the cooling electron model
is much flatter than the spectra for accelerating electron models
(Figure~\ref{fig:ntbrem_2}).
In this model, the electron population is rising for nonrelativistic
electrons ($N[p] \propto p^2$ for $p \ll 1$) and flat for relativistic
electrons ($N[p] \approx$ constant for $p \gg 1$; eq.~[\ref{eq:cool}]).
The nonrelativistic bremsstrahlung cross-section is roughly proportional
to $p^{-2}$ (eq.~[\ref{eq:bh}]), while the relativistic cross-section
increases slowly (logarithmically) with $p$
(Blumenthal \& Gould 1970).
Thus, the nonthermal bremsstrahlung emission in the cooling electron
model is dominated by the highest energy electrons for which
equation~(\ref{eq:cool}) applies, which we have taken to be electrons
with $p = p_u = 300$.
Even in the hard X-ray spectral band (say, 10 - 200 keV), most of the
emission is from electrons with energies of $\sim 150$ MeV.

The low frequency limit of the bremsstrahlung cross-section for very
relativistic electrons is
(Blumenthal \& Gould 1970).
\begin{equation} \label{eq:rel}
\frac{d \sigma ( p , \epsilon , Z )}{d \epsilon} \approx
\frac{32 \pi}{3} \, \frac{e^6}{m_e^2 c^5 h} \, \frac{Z^2}{\epsilon} \,
\ln \left( \frac{2 p^2 m_e c^2}{\epsilon} \right)
\, ,
\end{equation}
in the limit $ \epsilon \ll p m_e c^2 $ and $p \gg 1$.
As a result, the hard X-ray bremsstrahlung in the cooling electron
models is given approximately by
\begin{eqnarray}
L_\epsilon & \approx & \frac{64 \pi}{3} \, \frac{e^6}{m_e^2 c^4 h} \,
\left[ \sum n_Z Z ( Z + 1 ) \right] \,
\, N_o \, p_u \,
\ln \left( \frac{2 p_u^2 m_e c^2}{\epsilon} \right)
\nonumber \\
& \approx & \frac{32 \pi}{3} \, \frac{e^6}{m_e^2 c^4 h} \,
\left[ \sum n_Z Z ( Z + 1 ) \right] \,
\, N_{nt}^{tot} \,
\ln \left( \frac{2 p_u^2 m_e c^2}{\epsilon} \right)
\, . \nonumber \\
& & \label{eq:nt_cool}
\end{eqnarray}
Here, the factor of $Z(Z+1)$ includes the effects of electron-electron
bremsstrahlung, which is important at relativistic energies.
Thus, the hard X-ray emission is determined primarily by the total number
of nonthermal electrons, which are mainly at high energies in this model,
and the hard X-ray spectrum only increases logarithmically as the frequency
decreases.
This explains the flat spectrum of the HXR emission in the cooling electron
model.

\section{Emission by Higher Energy Electrons} \label{sec:harder}

The nonthermal electron populations discussed here represent an
intermediate phase between thermal electrons and highly relativistic
electrons.
In the accelerating populations, the particles producing HXR emission by
nonthermal bremsstrahlung are being accelerated to higher energies;
in the cooling electron model, the electrons are returning to the thermal
distribution from higher energies.
Thus, it is also useful to consider the emission produced by higher energy
electrons which are part of the same population.
For example, higher energy electrons
(with $p \approx \gamma \sim 300$) will produce extreme ultraviolet (EUV)
emission by inverse Compton (IC) scattering of cosmic microwave background
(CMB) photons.
Even higher energy electrons (with $p \approx \gamma \sim 10^4$)
will emit HXR radiation by the same process.
These higher energy electrons will also produce synchrotron radio emission
at a level determined by the magnetic field in the cluster.
The details of these emission processes and techniques to calculate the
emission are discussed in Sarazin (1999).
In Table~\ref{tab:harder}, we give the logarithm of the IC extreme ultraviolet
(EUV) luminosity, $\log L_{EUV}$ (65--245 eV),
and 
the logarithm of the IC hard X-ray luminosity,
$\log L_{HRX}$ (IC) in the band 10--200 keV.
These luminosities assume a cluster redshift of $z = 0$;
they scale as $(1 + z )^4$ if the band is assumed to be that which is
observed at zero redshift.
We also give the logarithm of the radio synchrotron power
$\log L_{\nu}$ (radio) at a wavelength of 92 cm ($\nu = 326$ MHz),
assuming an intracluster magnetic field of $B = 1$ $\mu$G.

%
%
\begin{table*}[htb]
\tabcaption{\hfil Emission from Higher Energy Electrons \label{tab:harder} \hfil}
\begin{center}
\begin{tabular}{lccccc}
\tableline
\tableline
&$\log L_{EUV}$&$\alpha_{*}$&$\log L_\nu$ (radio)&$\log L_{HXR}$ (IC)&
$\log L_{HXR}$ (NTB)\cr
Model&(ergs s$^{-1}$)&&(ergs)& (ergs s$^{-1}$)&(ergs s$^{-1}$)\cr
\tableline
Power-law, $\mu =2.0$ & 44.86 & $-$0.50 & 36.76 & 46.52 & 43.87 \cr
Power-law, $\mu =2.3$ & 44.04 & $-$0.65 & 35.47 & 45.30 & 43.75 \cr
Power-law, $\mu =2.6$ & 43.20 & $-$0.80 & 34.17 & 44.08 & 43.64 \cr
Power-law, $\mu =3.0$ & 42.09 & $-$1.00 & 32.43 & 42.45 & 43.53 \cr
Power-law, $\mu =3.5$ & 40.70 & $-$1.25 & 30.27 & 40.42 & 43.42 \cr
Power-law, $\mu =4.0$ & 39.30 & $-$1.50 & 28.11 & 38.41 & 43.34 \cr
Nonlinear, $p_c =0.3$ & 44.73 & $-$0.50 & 36.63 & 46.39 & 43.78 \cr
Nonlinear, $p_c =0.5$ & 44.56 & $-$0.50 & 36.46 & 46.22 & 43.68 \cr
Nonlinear, $p_c =1.0$ & 44.16 & $-$0.50 & 36.06 & 45.83 & 43.51 \cr
Cooling Electrons     & 47.15 & $-$1.47 &  ---  &  ---  & 43.37 \cr
\tableline
\end{tabular}
\end{center}
\end{table*}

The accelerating electron models are very close to power-laws at high
energies, and the IC and synchrotron spectra are also power-laws, with
a spectral index given by $\alpha_{*}$ in Table~\ref{tab:harder}.
Also, the radio powers given in Table~\ref{tab:harder} scale with
the intracluster magnetic field as $( B / 1~\mu$G$)^{1-\alpha_{*}}$.

The cooling electron model is expected to have a relatively flat
distribution up to an energy of $p \approx \gamma \sim 300$,
which is where the loss time of electrons in clusters is maximum.
The electron population is expected to drop rapidly to even higher
energies in a manner which depends on the past history of particle
acceleration in the cluster (Sarazin 1999).
We have adopted Model 11 from Sarazin (1999) for the higher energy
behavior of the cooling electron model.
This is a model which provides a reasonable approximation to the observed
EUV properties of clusters (Sarazin \& Lieu 1998).
Model 11 has a low-energy behavior which is well-approximated by
equation~(\ref{eq:cool}).
We have renormalized the results for Model 11 in Sarazin (1999) to
agree with the normalization of the cooling electron model in the
present paper.
This cooling electron model has an electron spectrum which is strongly concave
(decreasing rapidly at high energies), so the spectral index given in
Table~\ref{tab:harder} applies only to the EUV band.
The cooling electron model produces no significant radio synchrotron or
IC HXR emission.

In the accelerating electron models with the flatter spectra
(the power-law models with $\mu \la 2.7$ and the nonlinear shock models),
the IC hard X-ray luminosity of the extended electron population
significantly exceeds the nonthermal bremsstrahlung hard X-ray emission
by the same population.
For ease of comparison, the last column of Table~\ref{tab:harder} repeats
the values of the nonthermal bremsstrahlung hard X-ray luminosity
$L_{HXR}$ (NTB) from Table~\ref{tab:models}.
These models have nonthermal electron distributions similar to those
seen in the strong shocks of supernova remnants in our Galaxy.
Nonthermal bremsstrahlung is not interesting as a source for
hard X-ray radiation in these models, as more emission is produced by
IC scattering.
The radio powers of these models in Table~\ref{tab:harder} are also larger
than the diffuse radio emission observed in clusters with radio halos.
However, these radio luminosities could be decreased if the
intracluster magnetic field were very low, $B \ll 1$ $\mu$G.

Nonthermal bremsstrahlung is only interesting as a source of HXR
if the spectrum of accelerating electrons is steeper than is expected
for strong shocks, or if the electron spectrum cuts off or steepens
considerably at some energy $\ga$ 100 keV.
Earlier, we noted that steeper acceleration spectra might occur
in clusters because of the lower compression of intracluster shocks,
or because turbulent acceleration was important.
Table~\ref{tab:harder} shows that the models with steep power-law
electron distributions do produce more nonthermal bremsstrahlung
hard X-ray emission than IC HXR emission.
Thus, these models seem more directly applicable to clusters.
These models also have diffuse radio luminosities which are more consistent
with those observed in clusters, assuming $B \sim 1$ $\mu$G.

The observed EUV luminosities of rich clusters seem to lie in the
range $10^{43 - 45}$ ergs s$^{-1}$
(Lieu et al.\ 1996a,b;
Mittaz, Lieu, \& Lockman 1998).
Thus, the accelerating electron models with flatter spectra might also
explain this emission, if the same spectra extend up to
$\gamma \approx p \sim 300$.
However, the spectra would need to drop steeply at higher energies
to avoid producing too much HXR by IC.
The difficulty with the cooling electron model is the very large EUV
luminosity required by the same population which would produce a
significant HXR luminosity.
The required EUV luminosity is $\ga$$10^2$ higher than those of any
observed clusters.
The higher energy electrons must be present in the cooling electron model,
since they are the source of the lower energy electrons which produce the
HXR emission by nonthermal bremsstrahlung.

\section{Comparison to Observations of HXR from Clusters} \label{sec:observe}

Nonthermal hard X-ray tails have been detected in the spectra of the
Coma cluster
(Fusco-Femiano et al.\ 1999)
and the Abell~2199 cluster
(Kaastra et al.\ 1998, 1999) with
{\it BeppoSAX}.
In our numerical models, we assumed an
average thermal electron density of $n_e = 0.001$ cm$^{-3}$.
The NTB HXR luminosity in models for individual clusters is proportional to
$\int n_e n_{nt} \, d V$, where $n_e$ and $n_{nt}$ are the thermal and
nonthermal electron densities and $V$ is the volume.
We have normalized our models based on the total nonthermal
electron number, 
$N_{nt}^{tot} = \int n_{nt} d V$.
Thus, the NTB luminosity in models for individual clusters is proportional
to the average thermal electron density, defined as
\begin{equation} \label{eq:ne1}
\langle n_e \rangle \equiv
\frac{\int n_e n_{nt} \, d V}{\int n_{nt} \, d V} \, .
\end{equation}
Unfortunately, we generally do not have any detailed information on
the distribution of nonthermal particles in the cluster, particularly
the low energy particles which would produce HXR emission by NTB.
The {\it BeppoSAX} observations barely detect the HXR emission, and do not
give a detailed radial dependence.
(In Abell~2199, there is some evidence that the HXR emission is more broadly
distributed than the thermal emission [Kaastra et al.\ 1999].
In fact, this distribution is more easily explained if the HXR emission
is due to IC rather than NTB
[Sarazin \& Lieu 1998].)
We will assume arbitrarily that $n_{nt} \propto n_e$, so that
\begin{equation} \label{eq:ne2}
\langle n_e \rangle =
\frac{\int n_e^2 \, d V}{\int n_e \, d V} \, .
\end{equation}
Another problem is that the standard beta-model fits to the thermal
gas distributions in clusters do not give convergent masses at large
radii.
Thus, the integral in the denominator of equation~(\ref{eq:ne2}) must
be cutoff at some large radii.
We adopt the radius $r_{500}$, the radius at which the cluster overdensity
is equal to 500, as our standard radius.
Typically, this is of the order of the largest radius at which the thermal
gas distribution can be measured, and is very roughly one half of the
virial radius.
The total number of thermal or nonthermal electrons reported below are
the numbers within this radius.
We calculate the average density and total numbers of electrons using
the beta-model fits to the {\it ROSAT} surface brightness profiles of
clusters given in Mohr, Mathiesen, \& Evrard (1999).
All comparisons to data assume
$H_0 = 50$ km s$^{-1}$ Mpc$^{-1}$ and $q_0 = 0.5$,

\subsection{Coma Cluster} \label{sec:observe_coma}

In Coma, the best-fit power-law model gives a flux of about
$2.2 \times 10^{-11}$ ergs cm$^{-2}$ s$^{-1}$ in the 20-80 keV band,
which corresponds to a luminosity of about $5.1 \times 10^{43}$
ergs s$^{-1}$ (20-80 keV).
The beta-model fit to the {\it ROSAT} PSPC X-ray image gives
a central thermal electron density of $n_e (0) = 3.12 \times 10^{-3}$ cm$^{-3}$,
a core radius of 0.386 Mpc, and $\beta = 0.705$
(Mohr et al.\ 1999).
This implies that the total ICM mass is $2.23 \times 10^{14} \, M_\odot$,
and the total number of thermal electrons is
$N_{th}^{tot} = 2.29 \times 10^{71}$.
The average thermal electron density from equation~(\ref{eq:ne2}) is
$\langle n_e \rangle = 4.6 \times 10^{-4}$ cm$^{-3}$.

The spectral index of the HXR excess in Coma is very poorly determined;
the range is $-0.3 \le \alpha \le 1.5$ at the 90\% confidence level
(Fusco-Femiano et al.\ 1999).
This doesn't restrict the models in Table~\ref{tab:models} very dramatically;
only the steepest power-law models would be excluded.
Thus, most of the models in Table~\ref{tab:models} could fit the observed
HXR emission (with nonthermal electron populations which are $\sim$3\%
of the thermal electrons) if the nonthermal population is cut off at some
energy above 100 keV (so that no other emission is produced).
To further restrict the models, we must compare the emission from
higher energy electrons (\S~\ref{sec:harder}) to observations.
This requires that we assume that the nonthermal electron populations
extend to higher energies.
As noted in (\S~\ref{sec:harder}), we can then eliminate most of the
accelerating electron models;
all of the models with asymptotic power-laws flatter than $\mu \approx 2.7$
produce more HXR by IC than by bremsstrahlung
(Table~\ref{tab:harder}).

The observed radio flux of the Coma cluster halo is 3.2 Jy at 326 MHz
(Venturi, Giovannini, \& Feretti 1990).
The radio spectral index is uncertain.
The overall integrated spectral index in the radio is about
$\alpha = -1.34$
(Kim et al.\ 1990;
Giovannini et al.\ 1993).
There is some evidence that the spectral index steepens at high
frequencies, possibly due to losses by the higher energy electrons
(Giovannini et al.\ 1993),
but this is not certain
(Deiss et al.\ 1997).
At low frequencies, the radio spectral index may flatten to
$\alpha = - 0.96$
(Fusco-Femiano et al.\ 1999).

We will consider two power-law accelerating electron models for Coma,
whose properties are listed in Table~\ref{tab:observe}.
The first row in the Table gives the observed values of some of the
properties for Coma.
The second and third rows give the values for the models.
The values given are the total number of nonthermal electrons $N_{nt}^{tot}$,
the ratio of the number of nonthermal to thermal electrons (as a
percentage),
the normalization factor for the electron spectrum $N_o$,
the hard X-ray luminosity $L_{HXR}$ in the  20--80 keV band, and
the best-fit spectral index in this band $\alpha_{HXR}$,
the fraction of the model HXR emission in this band which is due to
IC rather than nonthermal bremsstrahlung $f_{IC}$,
the radio power at 326 MHz $L_\nu$ (radio),
the required cluster magnetic field $B$,
and
the EUV luminosity in the 65--245 eV band $L_{EUV}$.
The model values of $L_{HXR}$ and $L_\nu$ (radio) are in parentheses
because these values were used to set the overall normalization of the
model $N_o$ and the cluster magnetic field, so the agreement of these
values with observations is an assumption.

First, we consider a model assuming the average radio spectra index
of $\alpha = -1.34$, which implies a power-law exponent of $\mu = 3.68$.
With a spectrum this steep, essentially all of the HXR emission is due
to NTB rather than IC.
The spectral index of the HXR emission is steeper than is allowed by the
observations.
A relatively strong magnetic field ($\sim$4 $\mu$G) is required to
reproduce the radio flux of Coma at 326 MHz.
For the second model, we adopt the low frequency integrated radio spectral
index of $\alpha = - 0.96$ from
Fusco-Femiano et al.\ (1999), which implies an exponent of
$\mu = 2.92$ for the power-law electron distribution.
In this model, most of the HXR is due to nonthermal bremsstrahlung, but
there is still a nonnegligible contribution ($\sim$27\%) from IC scattering.
The HXR spectral index is fairly steep ($-1.19$), but consistent with
the observational limits.
The required magnetic field to reproduce the radio emission is fairly
small. 
(Note: Fusco-Femiano et al.\ [1999] adopt a more detailed description of
the radio and electron spectrum; if we adopt their method, the values of
the magnetic fields in our models are approximately doubled.)
These two models probably bracket the range of possible NTB models for
the HXR emission.
The required values of the magnetic field range from ones which are
a bit smaller than those implied by equipartition in the radio halo
($\sim$0.4 $\mu$G; Giovannini et al. 1993;
En{\ss}lin \& Biermann 1998) to larger ones which are
consistent with the strong fields implied by the Faraday rotation of
individual radio galaxies
($\sim$6 $\mu$G; Feretti et al.\ 1995).
In these models, the nonthermal electron population of Coma would be a
small fraction (2--4\%) of the thermal population.

In both of the models for Coma, the predicted EUV emission is weaker than
that observed.
Thus, another component of electrons would probably be required to explain
this emission.
Since the electrons which generate the EUV by IC have long lifetimes
(Sarazin \& Lieu 1998), it is possible that the EUV is produced by
an older population of electrons, while the NTB HRX emission and radio
emission are due to electrons currently being accelerated.

%
%
\begin{table*}[htb]
\tabcaption{\hfil Comparison to Observed Clusters
\label{tab:observe} \hfil}
\small
\begin{center}
\begin{tabular}{lcccccccccc}
\tableline
\tableline
&Observed&$N_{nt}^{tot}$&$N_{nt}^{tot}/N_{th}^{tot}$&
$N_o$&$L_{HXR}$ (20--80 keV)&$\alpha_{HXR}$&$f_{IC}$&
$L_\nu$ (radio)&$B$&$L_{EUV}$\cr
Cluster&Model&($10^{69}$)&(\%)&($10^{68}$)&($10^{43}$ ergs s$^{-1}$)&&(\%)&
($10^{31}$ ergs)&($\mu$G)&($10^{43}$ ergs s$^{-1}$)\cr
\tableline
Coma &Observed      &    &   &    & 5.1 &$-$1.5 - 0.3&        & 7.6 &   &
5.0\phn\phn\cr
     &PL $\mu =3.68$&9.28&4.0&8.97&(5.1)&$-$1.79     &\phn0.05&(7.6)&4.10&0.016\cr
     &PL $\mu =2.92$&4.62&2.0&8.22&(5.1)&$-$1.21     &27.1\phn&(7.6)&0.16&1.05\phn\cr
&&&&&&&&&&\cr
A2199&Observed            &    &   &    & 4.0 &$-$0.81         &   &$<$13&&$\sim$5\cr
     &PL cutoff $\mu =2.2$&2.56&3.2&5.55&(4.0)&$-$0.83         &\phn0.00&(0)&
---& (0) \cr
     &PL $\mu =2.9$       &2.54&3.2&6.01&(4.0)&$-$1.21         &23.6\phn&($<$13)&
$<$0.26&0.68\cr
    &PL $\mu =3.33$       &6.47&8.1&5.42&(4.0)&$-$1.50         &\phn0.80&($<$13)&$<$1.71&0.08\cr
\tableline
\end{tabular}
\end{center}
\end{table*}

\subsection{Abell~2199} \label{sec:observe_a2199}

{\it BeppoSAX} observations also indicate that there is a hard X-ray
tail in the cluster Abell~2199
(Kaastra et al.\ 1998, 1999).
If this hard tail is modeled with a power-law spectrum, it corresponds to
a total luminosity of $(1.30 \pm 0.32) \times 10^{44}$ ergs s$^{-1}$ in
the 0.1--100 keV band and a spectrum index of $\alpha =-0.81 \pm 0.25$
(Kaastra et al.\ 1999).
The equivalent luminosity in the 20--80 keV band is about
$4.0 \times 10^{43}$ ergs s$^{-1}$.
If only the hard X-ray PDS data in the band 10--100 keV is fitted,
the flux is about the same, but the power-law is considerably steeper,
$\alpha \approx -1.5$
((Kaastra et al.\ 1998).
This is qualitatively consistent with NTB models, in which the spectrum
flattens at low energies
(Fig.~\ref{fig:ntbrem_pl}).
The PDS spectral fit is not corrected for the cluster thermal emission, so
this spectral index is probably a lower limit to that of the nonthermal
hard tail.

The determination of the average thermal electron density in Abell~2199
is complicated by the presence of a cooling flow in this cluster.
Mohr et al.\ (1999) fit a separate component for the cooling flow.
Here, we will include both components, and calculate the average thermal
electron density without removing the cooling flow.
This gives $\langle n_e \rangle = 6.8 \times 10^{-4}$ cm$^{-3}$, and a total
population of thermal electrons of
$N_{th}^{tot} = 7.92 \times 10^{70}$.
The average electron density would be smaller by about 25\% if the cooling
flow component were removed.

We first consider a power-law accelerating electron model in which the HXR
emission is due to NTB and the spectral index agrees with the best-fit value
for the 0.1--100 keV range of $-$0.81.
The required exponent of the electron distribution is $\mu \approx 2.2$.
Unfortunately, if this model is extended to higher electron energies,
the IC HXR emission greatly exceeds that due to NTB.
(The EUV emission also exceeds that observed; see below.)
Thus, this model is not consistent, unless the electron distribution 
is cutoff at some energy above 100 keV.
Some properties of such a cutoff power-law model are shown in
the first model line for Abell~2199 in
Table~\ref{tab:observe}.
It reproduces the HXR flux and spectral index of Abell~2199 by design,
but has no predictive power.

One can produce the HXR tail in Abell~2199 with a power-law model
without a cutoff if the spectrum is sufficiently steep.
The second model line in Table~\ref{tab:observe} shows a model with
a power-law accelerating electron population with $\mu = 2.9$.
The last model line is a model with $\mu = 3.33$, which produces a
hard X-ray tail with a spectral index of $\alpha = -1.5$.
This is the spectral index derived from the PDS data alone, without
correction for thermal emission, which probably represents the
steepest allowable spectral index.
In this steep spectrum model, only a small fraction of the HXR emission
comes from IC, and the observed upper limit on cluster radio emission
can easily be satisfied without requiring a very small intracluster
magnetic field.
The models in Table~\ref{tab:observe} require nonthermal electron
populations which are 3--8\% of the thermal population.

Abell~2199 has been detected with {\it EUVE}
(Kaastra et al.\ 1999;
Lieu, Bonamente, \& Mittaz 1999).
None of the published papers give a flux or luminosity for the excess
EUV emission beyond the thermal emission from the ICM.
From the count rates and absorption, we crudely estimate that the
EUV luminosity is $\sim 5 \times 10^{43}$ ergs s$^{-1}$.
As was the case in Coma, the observed EUV flux is higher than that produced
by any model in which most of the HXR emission is due to NTB.
Apparently, a distinct population is needed to produce the observed EUV
emission.

\section{Conclusions} \label{sec:conclusion}

Recently, hard X-ray tails have been detected in the Coma and Abell~2199
clusters with {\it BeppoSAX}
(Fusco-Femiano et al.\ 1999;
Kaastra, Bleeker, \& Mewe 1998;
Kaastra et al.\ 1999).
The observed HXR emission is greater than that expected 
from the thermal X-ray emission from the hot ICM.
The HXR excesses have been fit by power-law spectra,
and have generally been interpreted as inverse Compton scattering
of CMB photons by relativistic electrons ($\sim$ GeV) in the cluster.
The Coma cluster has a diffuse radio halo, which also requires
similar high energy electrons.
However, the comparison of the HXR and radio emission in Coma implies a
rather small ICM magnetic field of $B \approx 0.16$ $\mu$G
(Fusco-Femiano et al.\ 1999).
Abell~2199 doesn't exhibit any diffuse radio emission
(Kempner \& Sarazin 1999).
If the observed HXR emission in this cluster were due to IC,
it would imply a very strong upper limit on the magnetic field
of
$\la 0.07$ $\mu$G
(Kempner \& Sarazin 1999).

These limits on the ICM magnetic field in Coma and Abell~2199 could
be avoided if all or part of the HXR emission were due to some other
mechanism.
One suggestion is that the HXR emission is nonthermal
bremsstrahlung (NTB) from suprathermal electrons with energies
of $\sim$ 10 -- 200 keV
(Kaastra et al.\ 1998;
En{\ss}lin et al.\ 1999).
These nonthermal electrons would form a population in excess of
the normal thermal gas which is the bulk of the ICM.
We have calculated nonthermal bremsstrahlung (NTB) models for the cluster
hard X-ray (HXR) tails.
In these models, the HXR emission is due to suprathermal electrons with
energies of $\sim$10--200 keV.
We considered models in which these transrelativistic suprathermal particles are
either the low energy end of a population of electrons which are being
accelerated to high energies by shocks or turbulence
(``accelerating electron'' models).
We considered both power-law momentum distributions, and a parameterized
form based on simulations of nonlinear shock acceleration.
We included models with steeper acceleration spectra than are usually found in
supernova remnants, because the ICM shocks have lower compressions than
supernova remnant shocks and because turbulent acceleration may be
important in clusters.
We also considered a model in which these electrons are the remnant of
an older nonthermal population which is losing energy and rejoining
the thermal distribution as a result of Coulomb interactions
(``cooling electron'' models).
The suprathermal populations were assumed to start at an electron
kinetic energy which is $3 kT$, where $T$ is the temperature of the
thermal intracluster medium (ICM).

The calculated nonthermal bremsstrahlung spectra flatten at low photon
energies because of the lack on low energy nonthermal particles.
The accelerating electron models had HXR spectra which were nearly
power-laws from $\sim$20--100 keV.
However, the spectra were brighter and flatter than those given by the
nonrelativistic bremsstrahlung cross-section because of transrelativistic
effects.
The HXR spectrum of the cooling electron model was very flat, and most
of the X-ray emission in the HXR energy range (10-100 keV) actually
arises from electrons with much higher energies ($\sim$ 100 MeV).

The nonthermal electron populations discussed here represent an
intermediate phase between thermal electrons and highly relativistic
electrons.
We have also calculated the inverse Compton (IC)
extreme ultraviolet (EUV), HXR, and radio synchrotron emission by the
extensions of the same populations to higher energies.
For accelerating electron models with power-law momentum spectra 
flatter than $\mu \la 2.7$ (i.e., those expected from strong shock
acceleration),
the IC HXR emission exceeded that due to NTB.
Thus, these models are only of interest if the electron population is
cut-off at some upper energy $\la$1 GeV.
Similarly, flat spectrum accelerating electron models
produced more radio synchrotron emission than is observed from cluster if
the ICM magnetic field is $B \ga 1$ $\mu$G.
The cooling electron model generated vastly too much EUV emission as
compared to the observations of clusters.
Thus, NTB models for the HXR emission are only interesting if the particle
spectra are steep and/or they cut-off or steepen at high energies.

We compared our NTB models to the observed HXR tails in Coma and Abell~2199.
The NTB models require a nonthermal electron population which contains
about 3\% of the number of electrons in the thermal ICM.
If the suprathermal electron population is cut-off at some energy
above 100 keV, then the models can easily fit the observed HXR fluxes
and spectral indices in both clusters.
The required electron spectra would be similar to those produced by
strong shock acceleration ($\mu \approx 2.3$).
A similar conclusion was reached for the Coma cluster by
En{\ss}lin et al.\ (1999).
If the cutoff were at an energy of $\sim$100 MeV, these models might also
explain the observed EUV emission.

For accelerating electron models without a cutoff, the electron spectrum
must be rather steep $\ga 2.9$ to avoid producing too much IC HXR
emission.
This results in NTB HXR spectra that are also steep, and only marginally
consistent with observations of the HXR spectrum in Abell~2199
and Coma or the radio spectrum in Coma.
These NTB models would allow values of the ICM magnetic field which are
more consistent with values derived from Faraday rotation or estimates based
on equipartition.
In general, the NTB models without a cutoff which fit the HXR emission
produce too little EUV emission to agree with the observations of clusters.
On the other hand, the electrons which produce the EUV emission have rather
long lifetimes in clusters, and might represent an older population,
while the NTB HXR and radio emission would be due to electrons currently
being accelerated.

We conclude that nonthermal bremsstrahlung is a viable explanation for
the HXR tails seen in clusters.  However, while NTB models alleviate
some of the problems with IC models, they face other unique problems.
A combination of NTB and IC models may be successful, but none of the
NTB models we have discussed can produce a single population of
particles which can explain all of the nonthermal emission from
clusters.

\acknowledgements
We thank Andrew Strong for providing his GALPROP code, which was used
to check the bremsstrahlung cross-sections.
This work was supported in part by NASA grants NAG 5-3057 and NAG 5-8390.

\end{document}